\documentstyle[multicol,prl,aps,graphics,amssymb]{revtex}
\begin{document}
\title{Superconductivity in molecular solids with Jahn-Teller phonons}
\author{J.E. Han,$^{1}$ O. Gunnarsson,$^{2}$, V.H. Crespi$^{(1)}$}
\address{${}^{(1)}$Department of Physics, The Pennsylvania State University,
University Park, PA 16802-6300}
\address{${}^{(2)}$ Max-Planck-Institut f\"ur Festk\"orperforschung 
D-70506 Stuttgart, Germany}

\maketitle

\begin{abstract}
We analyze fulleride superconductivity at experimental doping levels,
treating the electron-electron and electron-phonon interactions on an
equal footing, and establish the existence of novel physics which
helps explain the unusually high superconducting transition
temperatures in these systems.  The Jahn-Teller phonons 
create a local (intramolecular) pairing that is
surprisingly resistant to the Coulomb repulsion, despite the weakness
of retardation in these low-bandwidth systems. The requirement for
coherence throughout the solid to establish superconductivity then
yields a very strong doping dependence to T$_c$, one consistent with
experiment and much stronger than expected from standard Eliashberg
theory.
\end{abstract}

\begin{multicols}{2}
The discovery of superconductivity in alkali-doped C$_{60}$,
persisting up to unexpectedly high temperatures ($T_c=33$
K\cite{Tanigaki} or $T_c=40$ K\cite{Palstra}), raises interesting
questions about superconductivity in low-bandwidth molecular
solids. Superconductivity arises from an effective attractive
interaction between the electrons.  In conventional superconductors a
net attractive interaction survives, in spite of the strong Coulomb
repulsion, thanks to retardation effects\cite{Anderson}. However,
retardation is small for the fullerides\cite{Gertrud,Dierk}, since the
molecular vibration frequencies are comparable to the bare electron
bandwidth.  We show that the combination of molecular solid 
character and coupling
to Jahn-Teller phonons produces a local pairing, important for
superconductivity, which is {\it not} strongly suppressed by the Coulomb
repulsion.  In addition, the transition temperature depends
anomalously strongly on the doping level.  The superconducting
mechanism in fullerides therefore differs in important ways from that
of conventional superconductors.

Conventional superconductors are studied in the Migdal-Eliashberg
theory, assuming a band width $W$ much larger than a typical phonon
frequency $\omega_{ph}$.  For the fullerides, $\omega_{ph}\sim W$, so
the Eliashberg theory is of questionable accuracy. This failure of
Eliashberg theory is typically thought to lower $T_c$\cite{Freericks}
(although the opposite has also been argued\cite{Pietronero}). 
Metallic fullerides have very large, nonsaturating resistivities in 
the normal state\cite{Hebard,Xiang}  suggesting ``bad metal'' behavior
\cite{badmetal} which is also expected to
reduce $T_c$\cite{badmetal}. However, we find that
$T_c$ in the fullerides is {\it not} generally lower
than expected from Eliashberg theory due to an unusual cancellation of
countervailing effects. The violation of Eliashberg theory asserts
itself explicitly in a very strong doping dependence of $T_c$.

In A$_3$C$_{60}$ (A= K, Rb), the three-fold degenerate $t_{1u}$ level
is partly occupied and couples strongly to eight H$_g$ intramolecular
Jahn-Teller phonons. We capture the essential physics using a model
with one $t_{1u}$ level and one H$_g$ mode per molecule, with a
dimensionless electron-phonon coupling strength $\lambda$. We also
include the hopping between the molecules and the Coulomb repulsion
$U$ between two electrons on the same molecule\cite{Jong}. The model
explicitly includes Jahn-Teller coupling and places no implicit
restrictions on the ratio $\omega_{ph}/W$ or the value of $\lambda$.
We refer to this model as the $T\times h$ problem. To reveal the novel
effects of Jahn-Teller character, we compare this model to a
nondegenerate ($a$) or two-fold degenerate ($e$) level interacting
with a non-Jahn-Teller A$_g$ or two-fold degenerate ($E_g$) phonon,
i.e., $T\times a$, $E\times e$ and $A\times a$ problems, respectively.
Typical parameters are $\lambda \sim 0.5-1$, $\omega_{ph}/W\sim
0.1-0.25$ and $U/W\sim 1.5-2.5$\cite{rmp}.

We circumvent the limitations of Eliashberg theory by using the
dynamical mean-field theory (DMFT)\cite{DMFT} with a non-perturbative
Quantum Monte-Carlo (QMC) technique\cite{Scalapino}.  The electron
self-energy is assumed to be ${\bf q}$-independent, allowing a mapping
of the lattice problem onto an effective impurity problem.  We study
superconductivity by applying a perturbation creating electron pairs
and calculating the corresponding response function, i.e.\ the ${\bf
q}=0$ pairing susceptibility $\chi$. A divergence of $\chi$ below a
temperature $T_c$ signals the onset of superconductivity\cite{chi}. We
write
\begin{equation}\label{eq:chi}
\chi=(1-\chi_0\Gamma)^{-1}\chi_0,
\end{equation}
where $\chi_0$ is obtained from products of two fully dressed electron
Green functions describing the propagation of two electrons (holes),
which do not interact with each other.  Eqn. (\ref{eq:chi}) then
defines the effective interaction $\Gamma$.  We define a local
(intramolecular) pairing susceptibility
\begin{eqnarray}\label{eq:chiloc}
&&\chi^{\rm loc}(\tau_1,\tau_2,\tau_3,\tau_4)=  \\
&&-\sum_{mm^{'}}\langle c_{m\uparrow}^{\dagger}(\tau_1)
c_{m\downarrow}^{\dagger}(\tau_2) c_{m^{'}\downarrow}(\tau_3)
c_{m^{'}\uparrow}(\tau_4)\rangle,\nonumber 
\end{eqnarray}
where $\langle ...\rangle$ denotes a thermal average and $m$ labels
the $t_{1u}$ levels on one molecule.  Then
\begin{equation}\label{eq:chidmft}
\chi^{\rm loc}=(1-\chi_0^{\rm loc}\Gamma^{\rm loc})^{-1}\chi_0^{\rm loc}.
\end{equation}
$\chi^{\rm loc}$ and $\chi_0^{\rm loc}$ can be calculated within DMFT;
this defines the local interaction $\Gamma^{\rm loc}$.  $\Gamma
\approx \Gamma^{\rm loc}$ should be a rather good approximation, since
the interaction is dominated by intramolecular phonons and an
intramolecular Coulomb repulsion.  Since $\chi_0$ can be calculated
within DMFT, $\chi$ follows from Eqn. (\ref{eq:chi}).

Putting $\tau_1=\tau_2$, $\tau_3=\tau_4$ and taking the Fourier 
transform with respect to $\tau_1-\tau_3$ in the $T\to 0$ limit,
we obtain
\begin{equation}\label{eq:spectralrep}
\chi^{\rm loc}(i\omega_n)=\int_0^{\infty}\rho^{\rm loc}
(\varepsilon)/(\varepsilon-i\omega_n),
\end{equation}
where
\begin{eqnarray}\label{eq:spectral}
&&\rho^{\rm loc}(\varepsilon)=\sum_n|\langle n,N-2|\sum_m c_m{\uparrow}
c_m{\downarrow}|0,N\rangle|^2 \nonumber \\
&&\times  \delta(\varepsilon-E_0(N)+E_n(N-2)) +...
\end{eqnarray}
Here $|n,N\rangle$ is the $n$th excited state of the system with
$N$ electrons and the energy $E_n(N)$. The 
term shown describes the removal of an electron pair;
"$...$" indicates the addition of an electron pair.

\begin{figure}[bt]
{\rotatebox{-90}{\resizebox{!}{2.5in}{\includegraphics
{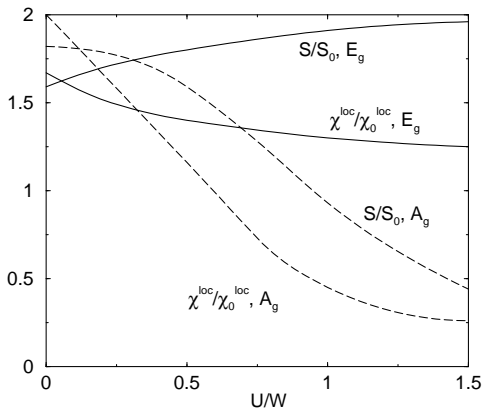}}}}
\caption[]{\label{fig:chi}
The ratios $S/S_0$ and $\chi/\chi_0$ drop as a function of $U/W$
for $\lambda=0.7$.  For the non-Jahn Teller model $A\times a$, 
these ratios drop rapidly as $U$ increases. In contrast, the 
pairing susceptibility for $E\times e$ is very resistant to 
increasing $U$. The results were obtained from exact diagonalization
for an impurity model with five host sites.
}
\end{figure}

The new physics can be best understood by examining a sum rule for the
spectral function $\rho^{\rm loc}(\varepsilon)$. In the simplest
Jahn-Teller case, $E\times e$,
\begin{equation}\label{eq:sumrule}
\int_0^{\infty}\rho^{\rm loc}(\varepsilon)d\varepsilon\equiv S = 4,
\end{equation}
for a half-filled band and in the limit of very large $U$ and very
small $\lambda$. In this limit the Jahn-Teller phonons produce local
singlets on the molecules,
\begin{equation}\label{eq:singlet}
{1\over \sqrt{2}}\sum_m c_{m\uparrow}^{\dagger}c_{m\downarrow}^{\dagger}
|vac\rangle,
\end{equation}
with pairing via the molecular quantum number $m$. Although the
Jahn-Teller effect competes with intermolecular hopping, a large U
reduces hopping, so that singlets form even for small $\lambda$. In
contrast, for $\rho_0^{\rm loc}(\varepsilon)$ the corresponding sum
rule gives only $S_0=1$. Since $\chi^{\rm loc}$ tends to be larger
than $\chi_0^{\rm loc}$, the effective interaction $\Gamma^{\rm loc}$
(Eq. (\ref{eq:chidmft})) tends to be attractive. The existence of
local singlets (Eq. (\ref{eq:singlet})) means that the probability
for removing or adding two electrons with the same $m$ quantum number
is very high. In contrast, for $\chi_0$ the electrons' $m$-quantum
numbers are independent and $\chi_0^{\rm loc}$ tends to be smaller.

As $U$ increases, spectral weight is shifted upwards in energy, which
tends to decrease $\chi^{\rm loc}$. However, this is partly
compensated by an increase of the integrated spectral weight $S$,
since the Jahn-Teller effect wins when hopping is reduced (see in
Fig. \ref{fig:chi} \cite{sumrule}).  Increasing $U$ therefore does not
rapidly eliminate a negative $\Gamma^{\rm loc}$, as one might have expected. In
contrast, for the $A\times a$ model the Coulomb repulsion $U$ and the
electron-phonon coupling directly compete, so $S$ (and therefore
$\Gamma^{\rm loc}$) drops quickly as $U$ is increased. These results
illustrate one important aspect of molecular solids with Jahn-Teller
phonons: counter-intuitively, Coulomb interactions can in certain 
respects actually {\it help} electron-phonon coupling. Capone {\it et
al.}\cite{Capone} have reached similar conclusions for A$_4$C$_{60}$
using a different approach.

Another important aspect is screening. The Coulomb interaction is
well-screened by the transfer of electrons between the
molecules\cite{Schluter,Gertrud,Erik}. Although this helps
superconductivity, it also normally implies an equally effective
screening of the electron-phonon interaction itself. However, since
the H$_g$ phonons do not shift the center of gravity of the electronic
levels, they cannot be efficiently screened by charge transfer on and
off the molecule\cite{Schluter,Erik}.

Both these effects are missing for A$_g$ phonons. A$_g$ phonons
furthermore tend to cause instabilities when coupled to a degenerate
level. Within a semiclassical approximation, a molecular solid with
$U=0$ becomes unstable when $\lambda \gtrsim 1/(2N)$, where $N$ is the
orbital degeneracy. A QMC calculation for a $T\times a$ model supports
this result, whereas in the $T\times h$ case the system stays metallic
for $\lambda \lesssim 1$ (and $U=0$). To be able to use a reasonably large
$\lambda$, we therefore study the $A \times a$ system below.

Although it is now clear that Jahn-Teller phonons can cause local
pairing, as described by $\chi^{\rm loc}/\chi^{\rm loc}_0$ and
$\Gamma^{\rm loc}$, superconductivity requires the formation of a
coherent state through the solid.  With a finite coherent metallic
weight at the chemical potential and an attractive interaction
$\Gamma$, the divergent unperturbed uniform pair propagator $\chi_0$
mediates a pairing instability towards forming the coherent
superconducting state.

\begin{figure}[bt]
{\rotatebox{-90}{\resizebox{!}{3.0in}{\includegraphics
{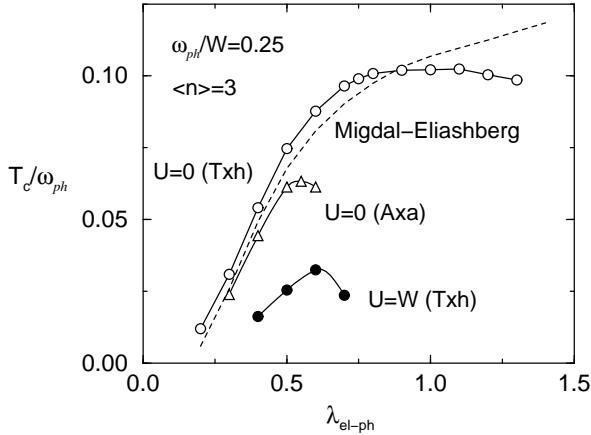}}}}
\caption[]{\label{fig:lambda}$T_c$ as a function of $\lambda$
according to Migdal-Eliashberg (dashed line) and DMFT
theories for the $T\times h$ ($\circ$) and 
$A\times a$ ({\tiny $\triangle$}) couplings at half-filling.
The parameters are $\omega_{ph}/W=0.25$ and $U=0$. 
The $T\times h$ results for $U=W$ ($\bullet$) are also shown.  }
\end{figure}

To obtain explicit results for the superconducting
transition, we use DMFT calculations solving the effective impurity
problem using QMC\cite{Hirsch,discretization}. We first discuss 
the case $U=0$.  Fig. \ref{fig:lambda} shows $T_c$ as a function of 
$\lambda$ according to the DMFT and Eliashberg\cite{self} theories.
The Eliashberg theory is expected to overestimate $T_c$ of doped C$_{60}$ 
both because of the violation\cite{Freericks} of Migdal's theorem and
because the mean-field Eliashberg equations are insufficient for a bad
metal\cite{badmetal}.  Surprisingly, in the $T \times h$ model for
$U=0$, the Eliashberg $T_c$ remains accurate even up to relatively
large values of $\lambda$. In contrast, for the $A\times a$ problem
the Eliashberg theory fails at smaller $\lambda$\cite{U=0}.  
The DMFT calculation shows a maximum $T_c$ at
$\lambda\sim 1$ due to a rapid drop of $\chi_0$ with increasing
$\lambda$ beyond this range (Eq. (\ref{eq:chi})). For small
$\lambda$, $\chi_0$ goes as $1/(1+\lambda)$, which renormalizes
$\lambda$ to $\lambda/(1+\lambda)$ in the McMillan
equation\cite{McMillan}.  For a larger $\lambda$, however, $\chi_0$
drops much faster in the DMFT than in the Eliashberg theory; the
formation of a coherent state is less efficient, since spectral weight
rapidly transfers away from the chemical potential as the system
approaches a metal-insulator transition.

\begin{figure}[bt]
{\rotatebox{-90}{\resizebox{!}{3.0in}{\includegraphics
{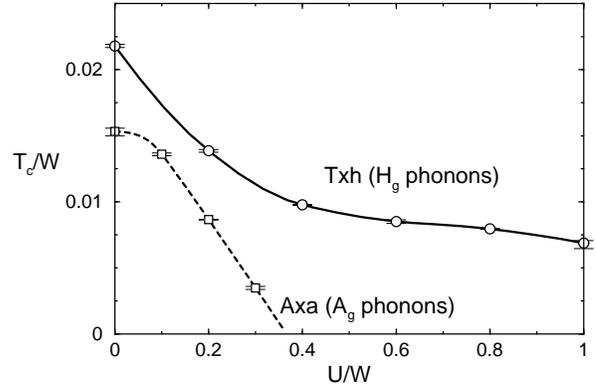}}}}
\caption[]{\label{fig:tc}$T_c$ as a function of $U$ for the 
$T\times h$ and $A\times a$ models for half-filling. The 
parameters are $\lambda=0.6$ and $\omega_{ph}/W=0.25$. The 
figure illustrates the important difference between H$_g$ 
and A$_g$ phonons.
}
\end{figure}

We next discuss finite $U$, connecting to Fig. \ref{fig:chi}. The
solid points of Fig. \ref{fig:lambda} show the overall reduction in
$T_c$ for finite $U$. Fig. \ref{fig:tc} shows $T_c$ as a function of
$U$ for the $T\times h$ and $A\times a$ models. For $A\times a$, $T_c$
drops quickly when $U$ increases, as expected.  However for
$T\times h$, $T_c$ is more resistant to increasing $U$\cite{TcMIT}.
This is consistent with the local pairing of Fig. \ref{fig:chi} and
illustrates the importance of treating explicitly the dynamic
interplay between Jahn-Teller phonons and electrons in molecular
systems with $W \sim \omega_{ph}$.

\begin{figure}[bt]
{\rotatebox{-90}{\resizebox{!}{3.0in}{\includegraphics
{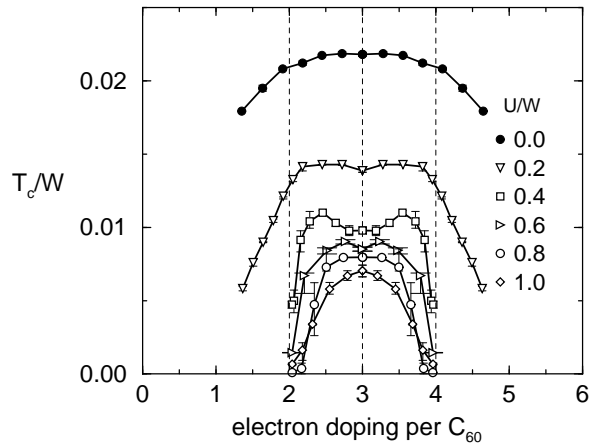}}}}
\caption[]{\label{fig:doping}$T_c$ as a function of doping $n$
for different values of $U$  for $T\times h$ coupling. The parameters
are $\omega_{ph}/W=0.25$ and $\lambda=0.6$. The figure illustrates the 
strong doping dependence for $U/W\ge 0.4$.
 }
\end{figure}
 
Experimentally, $T_c$ drops quickly in fullerides when the doping 
$n$ is reduced below three electrons per C$_{60}$ molecule\cite{Yildirim}. 
This cannot be explained within Eliashberg theory: reducing $n$ from 3
induces a slight {\it increase} of the density of states at the Fermi
energy\cite{Gelfand}, which should increase $\lambda$ and $T_c$.
This has been taken as evidence for an electron-electron mechanism of
superconductivity\cite{Kivelson}.  Fig. \ref{fig:doping} shows the
doping dependence of $T_c$ in DMFT.  For small $U$, $T_c$ drops slowly
until $n\sim 2$ or 4 and then starts dropping much faster:
$\Gamma^{\rm loc}$ drops rapidly here, probably because local pairing is
inefficient once the average number of electrons per molecule drops
below two.  For $U/W>0.4$, $T_c$ drops more quickly as $n=2$ is
approached. The system can gain a particular large Jahn-Teller energy
at $n=2$\cite{Tosatti1,Tosatti2}; this moves the system towards a
metal-insulator transition and shifts spectral weight in the
one-electron Green's function away from the chemical
potential\cite{Jong}. The shift in spectral weight rapidly reduces
$\chi_0$ and therefore $T_c$.  Thus the strong doping dependence can
be explained within an electron-phonon mechanism, and there is no need
to assume an electronic mechanism.

For conventional superconductors, retardation is important in reducing
the effects of the strong Coulomb repulsion\cite{Anderson}. For the
fullerides, the retardation effects are small, since $W\sim
\omega_{ph}$\cite{Gertrud,Dierk}. Local pairing and screening are
therefore crucial in reducing the effects of the Coulomb repulsion for
the fullerides. An increasing Coulomb interaction does not much damage 
superconductivity, since the concomitant reduction in hopping favors 
the Jahn-Teller pairing. This leads to new physics in these strongly 
correlated low-bandwidth molecular solids. The importance of local 
pairing is consistent with the short coherence length, which is only 
about three times the C$_{60}$-C$_{60}$ separation\cite{Holczer,Hou}.   

OG and JEH thank the Max-Planck Forschungspreis for support.  JEH and
VHC acknowledge the US National Science Foundation under grant
DMR-9876232, the NPACI, and the Packard Foundation.

\end{multicols}

\begin{thebibliography}{*}

\bibitem{Tanigaki}K. Tanigaki {\it et al.}, Nature {\bf 352}, 222 (1991).

\bibitem{Palstra}T.T.M. Palstra {\it et al.}, Solid State 
Commun. {\bf 93}, 327 (1995).

\bibitem{Anderson} P. Morel, and P.W. Anderson, Phys. Rev. 
{\bf 125}, 1263 (1962).

\bibitem{Gertrud}
O. Gunnarsson and G. Zwicknagl, Phys. Rev. Lett. {\bf 69}, 957 (1992).

\bibitem{Dierk}
O. Gunnarsson, D. Rainer, and G. Zwicknagl, Int. J. Mod. Phys. B {\bf 6},
3993 (1992).


\bibitem{Freericks}J.K. Freericks, Phys. Rev. B {\bf 50}, 403 (1994).

\bibitem{Pietronero}L. Pietronero, S. Str\"assler, and C. Grimaldi, 
Phys. Rev. B {\bf 52}, 10516 (1995). 

\bibitem{Hebard}A.F. Hebard {\it et al.},                     
Phys. Rev. B {\bf 48}, 9945 (1993).

\bibitem{Xiang}X.-D. Xiang {\it et al.}, Science {\bf 256},
1190 (1992).

\bibitem{badmetal}V.J. Emery and S.A. Kivelson, Phys. Rev. Lett.
{\bf 74}, 3253 (1995).

\bibitem{Jong}J.E. Han, E. Koch, and O. Gunnarsson, 
Phys. Rev. Lett. {\bf 84}, 1276 (2000).

\bibitem{rmp}O. Gunnarsson, Rev. Mod. Phys. {\bf 69}, 575
(1997).

\bibitem{DMFT}A. Georges, G. Kotliar, W. Krauth, M.J. Rozenberg,
Rev. Mod. Phys. {\bf 68}, 13 (1996).

\bibitem{Scalapino}R. Blankenbecler, D.J. Scalapino, and R.L. Sugar,
Phys. Rev. D {\bf 24}, 2278 (1981). 

\bibitem{chi}C.S. Owen and D.J. Scalapino, Physica {\bf 55}, 691
(1971).

\bibitem{sumrule} In the limit of both $U$ and $\lambda$ large, which 
is more appropriate to Fig. \ref{fig:chi}, the sum rule 
(\ref{eq:sumrule}) is reduced to $S=2$, since the states 
($c_{1\uparrow}^{\dagger}c_{2\downarrow}^{\dagger}|{\rm boson \ state}\rangle$) 
coupling the different components of the singlet (\ref{eq:singlet}) 
then have a substantial weight in the ground-state.  

\bibitem{Capone}M. Capone, M. Fabrizio, C. Castellani and
E. Tosatti, Science {\bf 296}, 2364 (2002).

\bibitem{Schluter}M. Schl\"uter {\it et al.},                        
J. Phys. Chem. Solids {\bf 53}, 1473 (1992).

\bibitem{Erik}E. Koch, O. Gunnarsson and R.M. Martin,
Phys. Rev. Lett. {\bf 83}, 620 (1999).
 
\bibitem{Hirsch}R. M. Fye and J. E. Hirsch, Phys. Rev. B
 {\bf 38} 433 (1988).

\bibitem{discretization}
Near Mott transition, the discretization of imaginary time 
in QMC technique becomes problematic,  {\it eg.} nearly 20 \% 
overestimate of $T_c$ at $U/W=1$ $\lambda=0.6$.

\bibitem{self} The Eliashberg calculation used a self-consistent 
phonon Green's function to lowest order in the phonon self-energy.

\bibitem{U=0} Near $\lambda\gtrsim 0.6$, $A\times a$ model becomes 
unstable due to the charge fluctuation, as pointed out earlier.


\bibitem{Yildirim}T. Yildirim {\it et al.},                           
Phys. Rev. Lett. {\bf 77}, 167 (1996).

\bibitem{Gelfand}M.P. Gelfand and J.P. Lu, Phys. Rev.
Lett. {\bf 68}, 1050 (1992). 

\bibitem{McMillan}W.L. McMillan, Phys. Rev. {\bf 167}, 331 (1968).

\bibitem{TcMIT}
The critical $U$ for the metal-insulator transition is
$U_c/W\approx 1.5$ for $\lambda=0.6$\cite{Jong}. Although not shown here, the
superconductivity is expected to persist up to $U=U_c$.             

\bibitem{Holczer}K. Holczer {\it et al.},                          
Phys. Rev. Lett. {\bf 67}, 271 (1991).

\bibitem{Hou}J.G. Hou {\it et al.}, Solid State Commun. {\bf 86}, 643 
(1993).

\bibitem{Kivelson}
S. Chakravarty, M. Gelfand, and S. Kivelson, Science {\bf 254},
970 (1991).


\bibitem{Tosatti1}A. Auerbach, N. Manini, and E. Tosatti,
Phys. Rev. B {\bf 49}, 12998 (1994). 

\bibitem{Tosatti2}N. Manini, E. Tosatti,
and A. Auerbach, Phys. Rev. B {\bf 49}, 13008 (1994).



\end{thebibliography}
\end{document}